There is No Violation of Conservation Laws in Quantum Measurement

Dr. Jacques Mallah

(updated 7/2/2007)

jackmallah@yahoo.com

Abstract:

For some ideal quantum measurements, conservation laws would seem to be violated systematically. It is argued that the intrinsically non-"ideal" nature of quantum measurements rescues the conservation laws.

1.   Introduction

In the standard description of measurement on a quantum mechanical system it is far from obvious that conservation laws are obeyed. As described below for the example of the Stern-Gerlach device (SGD) for measuring spin [1], conservation of angular momentum would be seemingly violated [2] in a sense described below if an ideal textbook measurement of spin could be performed.

Consider an ideal measurement of the Z-component of the spin of an incoming spin-½ particle by a SGD. This will be described for the cases in which the incoming particle is already in an eigenstate of the Z-component of spin with the following notation:

$$|+z\rangle \otimes |ready\rangle \rightarrow |u\rangle$$

$$|-z\rangle \otimes |ready\rangle \rightarrow |d\rangle$$

where

|+z> is the spin state of the incoming particle that is parallel to the Z direction

|-z> is the spin state of the incoming particle that is opposite to the Z direction

|ready> is the initial state of the SGD and associated lab (including any postdocs and students) ready to measure the spin

|u> is the state of the particle, SGD and lab after the SGD found a spin parallel to the Z direction

|d> is the state after the SGD found a spin opposite to the Z direction

➔ indicates the passage of time

The orbital angular momentum of the incoming particle will be initially identical for both cases (and is included in the |ready> ket). States |u> and |d> implicitly depend on the time after the measurement. It is not assumed that the spin of the measured particle remains constant after the measurement, which is why the |u> and |d> states are taken to include that degree of freedom as well as any others.

It is implicitly assumed above that the SGD (and lab) has a sufficiently certain angular orientation, due to its macroscopic nature, that any error terms arising from the initial uncertainty in the direction of the SGD can be ignored. It will be shown below that if that assumption were true, conservation of angular momentum would appear to fail.

Consider an incoming particle that has its spin at some other angle. This is described by a superposition of the above cases:

$$(a\ |+z> + b\ |-z>) \otimes |ready> \ ➔ \ a\ |u> + b\ |d> = |final>$$

(Unitary quantum mechanics (also known as the many worlds interpretation [3,4]) will be assumed to hold, but this paper is largely independent of the interpretation of quantum mechanics.)

The expectation value of the angular momentum of the |ready> state, <ready|$J_{SG}$|ready>, will be assumed to be zero to simplify the analysis. The expectation value of the angular momentum of the |+z>⊗|ready> state must equal that of the |u> state, and the expectation value of the |−z>⊗|ready> state must equal that of the |d> state, because total angular momentum is conserved. These must be equal to the spin of the incoming particle, ½ ℏ in the +Z or −Z direction, respectively:

$$<u|J|u> = (0, 0, ½ ℏ)$$

$$<d|J|d> = (0, 0, -½ ℏ)$$

For the superposition, the initial expectation value for angular momentum must be (½ ℏ) $\mathbf{u}_s$, where $\mathbf{u}_s$ is the unit vector in the direction of the spin of the incoming particle. Explicitly,

$$\mathbf{u}_s = (u_x, u_y, u_z) = (2\,\text{Re}\,(a^* b),\ 2\,\text{Im}\,(a^* b),\ a^* a - b^* b)$$

where the normalization condition $a^* a + b^* b = 1$ is assumed.

Now consider the expectation value of the angular momentum for the |final> state:

$$<\text{final}|\,J\,|\text{final}> = <J> =$$

$$(a^* <u| + b^* <d|)\,J\,(a\,|u> + b\,|d>) =$$

$$a^* a\,<u|J|u> + 2\,\text{Re}\,[a^* b\,<u|J|d>] + b^* b\,<d|J|d>$$

By overall conservation of angular momentum, this must be equal to the initial expectation value, ½ ℏ $\mathbf{u}_s$:

$$\langle J_x \rangle = \tfrac{1}{2}\hbar\,(2\,\mathrm{Re}\,(a^*b)) =$$

$$a^*a\,\langle u|J_x|u\rangle + 2\,\mathrm{Re}\,[a^*b\,\langle u|J_x|d\rangle] + b^*b\,\langle d|J_x|d\rangle$$

$$\langle J_y \rangle = \tfrac{1}{2}\hbar\,(2\,\mathrm{Im}\,(a^*b)) =$$

$$a^*a\,\langle u|J_y|u\rangle + 2\,\mathrm{Re}\,[a^*b\,\langle u|J_y|d\rangle] + b^*b\,\langle d|J_y|d\rangle$$

$$\langle J_z \rangle = \tfrac{1}{2}\hbar\,(a^*a - b^*b) =$$

$$a^*a\,\langle u|J_z|u\rangle + 2\,\mathrm{Re}\,[a^*b\,\langle u|J_z|d\rangle] + b^*b\,\langle d|J_z|d\rangle$$

By matching coefficients for a and b it can be concluded that

$$\langle u|J_x|d\rangle = \tfrac{1}{2}\hbar$$

$$\langle u|J_y|d\rangle = -i\,\tfrac{1}{2}\hbar$$

$$\langle u|J_z|d\rangle = 0$$

2.  Macroscopic separation

The brackets derived in section 1 under the assumptions given have interesting implications when the fact that |u> and |d> represent macroscopically decoherent branches of the wavefunction is kept in mind. A more careful look at this fact, however, will reveal that those brackets must be incorrect.

I.  Type I effective violation: Correlation

Assume, for now, the above results. For angular momentum in the direction of the measurement, the Z direction, the expectation value $\langle J_z \rangle$ is conserved overall, but in each macroscopically decoherent branch the average value of $J_z$ (for example, $\langle u|J_z|u\rangle$ for the "up" branch) will appear to be either $+\frac{1}{2}\hbar$ or $-\frac{1}{2}\hbar$, depending on which branch it is. This is no surprise; in each branch, it seems as though the other branch does not exist.

There is therefore in each branch the appearance (illusion) of the violation of conservation of $\langle J_z \rangle$ for the whole system, which can be called an "effective" (or seeming) violation. (Here, "effective" implies "for practical purposes". If "collapse of the wavefunction" were assumed, this would be assumed to be a real violation.)

$$|\text{final}\rangle = a\,|u\rangle + b\,|d\rangle$$

$$\langle u|J_z|u\rangle = \tfrac{1}{2}\hbar$$

$$\langle d|J_z|d\rangle = -\tfrac{1}{2}\hbar$$

$$\langle J_z\rangle_{\text{initial}} = \langle J_z\rangle_{\text{final}} = \langle\text{final}|J_z|\text{final}\rangle$$

$$= a^*a\,\langle u|J_z|u\rangle + b^*b\,\langle d|J_z|d\rangle$$

This type of "violation" of a conservation law, in which the cross terms such as $\langle u|J_z|d\rangle$ are not involved, will be called a type I violation.

In general, a type I violation occurs for an operator F that commutes with the Hamiltonian if

$$\langle F\rangle_{\text{initial}} = \langle F\rangle_{\text{final}}$$

$$= \Sigma_i\,(\langle i|\,c_i^*)\,F\,(c_i\,|i\rangle) = \Sigma_i\,c_i^*\,c_i\,\langle i|\,F\,|i\rangle$$

where each $|i\rangle$ represents a macroscopically decoherent "branch" of the wavefunction, and $c_i$ is its coefficient in the final superposition, and if the terms <i|F|i> are not all equal. In other words, if the cross terms between "branches" vanish, then the average value of <F> on all branches weighted by the standard measure [3,4] for each branch is equal to the initial expectation value of F, and there is a type I violation (unless the <i|F|i> are equal in which case there is no violation).

If one were to measure the total angular momentum of the lab in the Z direction after the experiment, the value obtained would be correlated with the outcome of the SG experiment that happened within the lab; but the experimentally obtained total angular momentum in the Z direction would, on average, be equal to what it would have been if the angular momentum in the Z direction of the |initial> state were measured instead.

II. Type II effective violation: Cross-term Storage

Going back to the results in part 1, for angular momentum in a perpendicular direction to Z, such as the X direction, there is a systematic effective violation that is always in a predictable direction. $<u|J_x|u> = 0$ and $<d|J_x|d> = 0$, so in each branch it will appear that the angular momentum carried by the spin of an incoming particle that was polarized in the X direction has vanished. The "missing" angular momentum has been "stored" in the cross-term, $<u|J_x|d>$, which is not detectable because there is no way in practice for an observer in one branch to detect the other branch, or even for an outside observer to detect both branches simultaneously. This type of violation, discussed in [2], will be called a type II violation.

If this is so, and the experiment is repeated N times, the violation is proportional to N. For example, consider a lab within a satellite. A steady stream of particles with their spins all polarized to be in the +X direction – produced by angular-momentum-conserving processes not involving 'measurement' – could be sent into the SGD. Upon measurement in the Z direction, this angular momentum in the +X direction would appear to vanish. By this means the total angular momentum of the satellite would steadily change with time, gaining angular momentum in the -X direction.

A type II violation is a problem because it's too good to be true: for all practical purposes, it could be used to violate conservation laws in a desired direction, which seems very unlikely. If conservation laws could be violated in a predictable way, the effect would most likely have been noticed in simpler systems than a SGD, with interaction with the environment providing the needed measurement-like decoherence.

A closer look at the cross-term, $<u|J_x|d> = ½ \hbar$ = constant, leaves no doubt that the analysis in section 1 must be in error. One reason is that (contrary to the assumption of O'Flanagan [2]), the angular momentum operator [1] is of the form $\mathbf{J} = \mathbf{S} \otimes 1 + 1 \otimes \mathbf{L}_{SG}$, where $\mathbf{S}$ is the spin of the incoming particle and $\mathbf{L}_{SG}$ is the angular momentum of the detector. For an operator of this form, if the spin of the measured particle stays fully correlated with the measurement outcome (which it could), the term $<u|\mathbf{J}|d>$ must vanish as shown in [2].

However, there is a more general and important reason that such a term must vanish. Macroscopic distinguishability of branches, which I will call macroscopic decoherence, doesn't just mean $<u|d>$ is small; it means that $|u>$ and $|d>$ could be as completely unlike each other as a world with a live cat vs. a world with a dead cat. The corresponding wavefunctions are large in <u>completely different regions</u> of 'classical configuration' space, and these become more and more different as time goes on. For an n particle system

$$<u|F|d> = \Sigma_{spins} \int dx^{3n} \, \Psi_u^* \, F \, \Psi_d$$

The integral must be nearly zero if the configurations (spins, positions) in which the functions $\Psi_u^*$ and $F \Psi_d$ are non-negligible are macroscopically separated. For example, in the "up" branch the postdoc could go to the cafeteria and eat a sandwich, while in the "down" branch he could be instructed to continue the experiments. These wavefunctions will not have a lot of configuration-space overlap, and applying the operator F to $\Psi_d$ to create the function $F\Psi_d$ will still not result in a function that overlaps with $\Psi_u^*$. Macroscopic distinguishability is a much stronger statement than mere orthogonality of the functions.

Only if F were a very complicated operator tailored for the specific conditions would it be plausible that <u|F|d> could be non-negligible. Therefore, <u|F|d> = 0 holds to an excellent approximation for any commonly encountered operator F, including angular momentum, and a type II violation (cross-term storage) could never occur.

3. The SGD as a quantum system

The apparent paradox can be resolved by treating the measuring apparatus as a fully quantum system. The approximation made in part 1 that the SGD could be assumed to have a definite angular orientation is inadequate if the angular momentum of the SGD itself is to be considered. In fact, any system with a definite orientation would have an infinitely uncertain angular momentum.

The uncertainty in the orientation of the SGD implies that it may register a "down" result even if the incoming spin was purely in the positive Z direction, and may register an "up" result for an incoming spin that was purely in the negative Z direction:

$$|+z>\otimes|ready> \rightarrow C|u> + D|d'> = |p>$$

$$|-z>\otimes|ready> \rightarrow E|d> + F|u'> = |q>$$

The "error" state |d'> is different from the "ordinary" down state |d>, and |u> ≠ |u'>, although they are not macroscopically distinguishable. For

convenience, C, D, E, and F will be taken to be real numbers by absorbing the phases into the kets.  In practice, there would also be other sources of error, but that is not required for this analysis and will be neglected.

Since the SGD is a large object of fairly well defined direction, C ≈ 1 and E ≈ 1, while D and F are small. The magnitude of D and F can be estimated by assuming that the uncertainty in the orientation of the SGD, Δθ, is the minimum consistent with the uncertainty in angular momentum ΔL for a typical member of the thermal ensemble appropriate to the device, so that Δθ ≈ ℏ / ΔL.

Using $(\Delta L)^2 \approx \langle L^2 \rangle_{thermal} \approx I k T$, where I is the moment of inertia, k = 1.3807 x 10$^{-23}$ J/K is Boltzmann's constant and T is the temperature, and assuming a moment of inertia I ≈ .01 kg-m$^2$, and a temperature T of 300 K, this implies I k T ≈ 4 x 10$^{-23}$ kg²m⁴/s² or ΔL ≈ 6 x 10$^{-12}$ kg-m²/s:

$$\Delta\theta \approx \hbar / (I k T)^{1/2} \approx 10^{-22} \text{ radians}$$

Since the amplitude of an 'error' due to incorrect orientation of the SGD is equal to the average of the sine of half of the angle between the axis of the SGD and the incoming particle, and it is a small angle, E ≈ F ≈ Δθ / 2. More crudely one may say E ≈ F ≈ Δθ.  The small size of Δθ may make it seem negligible, but that will not matter; what is important is ΔL ≈ ℏ / Δθ remains reasonably small, and not infinite as would be the case if Δθ = 0.

In the general case,

a |+z⟩|ready⟩ + b |−z⟩|ready⟩ ➔   a |p⟩ + b |q⟩ = |final⟩

and the two macroscopically distinct states are

N |up⟩ = a C |u⟩ + b F |u'⟩

$$M \,|dn\rangle = b\,E\,|d\rangle + a\,D\,|d'\rangle$$

where N and M are coefficients chosen so that $\langle up|up\rangle = \langle dn|dn\rangle = 1$. F and D are small and $C \approx E \approx 1$, so $N \approx a$ and $M \approx b$.

Next, consider conservation of overall angular momentum for the same situation as before. Although there will be more terms in the equations, cross terms involving different macroscopically distinct states (such as $\langle u|J_x|d\rangle$ for example) will be taken to equal zero, as this should be an excellent approximation as discussed in the previous section.

For an incoming spin in the $|+z\rangle$ direction this gives

$$\langle p|\,\mathbf{J}\,|p\rangle = (0,\,0,\,\tfrac{1}{2}\hbar)$$

$$= (C\,\langle u| + D\,\langle d'|)\,\mathbf{J}\,(C\,|u\rangle + D\,|d'\rangle)$$

$$= C^2\,\langle u|\,\mathbf{J}\,|u\rangle + D^2\,\langle d'|\,\mathbf{J}\,|d'\rangle$$

Since $D \approx \Delta\theta \ll 1$, $\langle u|J|u\rangle = \langle p|J|p\rangle$ to a very good approximation.

Similarly, for an incoming spin in the $|-z\rangle$ direction

$$\langle q|\,\mathbf{J}\,|q\rangle = (0,\,0,\,-\tfrac{1}{2}\hbar)$$

$$= (E\,\langle d| + F\,\langle u'|)\,\mathbf{J}\,(E\,|d\rangle + F\,|u'\rangle)$$

$$= E^2\,\langle d|\,\mathbf{J}\,|d\rangle + F^2\,\langle u'|\,\mathbf{J}\,|u'\rangle$$

For the expectation value of angular momentum in the general case, the additional terms now make it possible to conserve the angular momentum expectation value while setting cross terms between macroscopically distinct states to zero:

$$\langle\text{initial}|\ J\ |\text{initial}\rangle$$

$$= \tfrac{1}{2}\hbar\ (2\ \text{Re}\ (a^*b),\ 2\ \text{Im}\ (a^*b),\ a^*a - b^*b) =$$

$$\langle\text{final}|\ J\ |\text{final}\rangle = N^*N\ \langle\text{up}|\ J\ |\text{up}\rangle + M^*M\ \langle\text{dn}|\ J\ |\text{dn}\rangle =$$

$a^*a\ C^2\ \langle u|J|u\rangle + 2\ C\ F\ \text{Re}\ [a^*b\ \langle u|J|u'\rangle] + b^*b\ F^2\ \langle u'|J|u'\rangle +$
$b^*b\ E^2\ \langle d|J|d\rangle + 2\ E\ D\ \text{Re}\ [b^*a\ \langle d|J|d'\rangle] + a^*a\ D^2\ \langle d'|J|d'\rangle$

Matching coefficients,

$$C\ F\ \langle u|J_x|u'\rangle + E\ D\ \langle d|J_x|d'\rangle = \tfrac{1}{2}\hbar$$

$$C\ F\ \langle u|J_y|u'\rangle + E\ D\ \langle d|J_y|d'\rangle = -i\ \tfrac{1}{2}\hbar$$

$$C\ F\ \langle u|J_z|u'\rangle + E\ D\ \langle d|J_z|d'\rangle = 0$$

The states $|u\rangle$ and $|u'\rangle$ must be different because otherwise the above brackets would be impossible, and likewise for the down states. Since $C \approx E \approx 1$ and $D \approx F \approx \Delta\theta$, this implies that $\langle u|J_x|u'\rangle$ is of order $\hbar\ /\ \Delta\theta$, or in other words, is of order $\Delta L$ which is the uncertainty in angular momentum of the SGD within the $|\text{initial}\rangle$ state.

    It is remarkable that the effect of the 'error' terms on the angular momentum expectation value is of order $\hbar$, which is the same order of magnitude as the effect of the normal terms for this spin measurement, despite the much smaller amplitude of the 'error' terms. However, it is not unreasonable for an angular momentum bracket to be of order $\Delta L$, which is the uncertainty in the device's angular momentum.

    The relationship obtained with macroscopically distinct cross terms set to zero

$$\langle \text{initial}|J|\text{initial}\rangle = \langle \text{final}|J|\text{final}\rangle = N^*N \langle \text{up}|J|\text{up}\rangle + M^*M \langle \text{dn}|J|\text{dn}\rangle$$

is characteristic of Type I violation and not of Type II violation.

The initial angular momentum in directions perpendicular to the measurement direction therefore need no longer be seen as having 'vanished' for an observer, who sees only one of the two macroscopically distinguishable outcomes. The state he is part of such as

$$|\text{up}\rangle = a\, C/N\, |u\rangle + b\, F/N\, |u'\rangle$$

has a microscopic dependence on the original angle of the incoming particle's spin, and this is enough to allow the "missing" angular momentum to be stored within such states with no need for macroscopically-distinct-cross-term storage.

4. The final nail: Type I violation revisited

Suppose the satellite that contains the SGD has an unusual run of luck, and while measuring in the Z-direction particles that were initially polarized in the +X direction, many particles in a row are found to have spins in the +Z direction and few are in the -Z direction. Does this freak run of luck lead to macroscopic (apparent) violation of the conservation law from a Type I violation?

The scalar square of the total angular momentum, $J^2$, is a conserved quantity just as the total angular momentum vector is. But if the experiment in the satellite is continued, and 'lucky streaks' produce apparent violations of the conservation law for angular momentum, it would now seem that the apparent expectation value of $J^2$ would be certain to grow with time, because lucky streaks are sure to happen eventually and $J^2$ would increase regardless of their direction.

This kind of systematic growth seems implausible and at odds with the on-average expectation-value respecting properties that Type I violations seemed to have. Can this violation be eliminated, restoring apparent conservation laws?

If the source of the incoming particles is external to the satellite, there is no problem, because the incoming stream of particles has an intrinsic uncertainty in its total Z-direction angular momentum, and it serves as an external source of angular momentum that naturally would be expected to increase the total $J^2$ within the satellite.

Just as it was necessary to take the uncertainty in orientation of the SGD into account, it is equally true that the source of our incoming particles must have its own uncertainty in orientation. This must be taken into account when the source itself is within the system under consideration, in this case the satellite.

As a result of the angular uncertainty of the source, the incoming particles can not be pure states, but are entangled in orientation with the source, even though the angles by which they could deviate from the desired direction are microscopically small. This provides a way for the results of the SGD's measurements to become correlated with the angular momentum of the source; and as shown in the above analysis of the SGD, even microscopically small corrections to the wavefunction can make the right kind of difference.

The implication is that the overall angular momentum of the whole system – in this case, the enclosed satellite – will still appear to be conserved regardless of the results of the SGD measurements even for 'lucky streaks'. The satellite will need to rely on external torques and thrusters to change its angular momentum.

5. Conclusion

In standard formulations of thought experiments involving measurement, such as the Stern-Gerlach experiment, an apparent systematic effective violation of conservation laws appears. This is due to a failure to properly include the uncertainty in the conserved quantities possessed by the macroscopic measuring device in question. Proper modeling of the experiment implies the existence of small 'error' terms which permit the final macroscopic states to absorb the quantity. The specific example of a Stern-Gerlach device was used above, but it has been shown [6] that no measurement of something that does not commute with every conserved quantity can be 100% reliable, so that such 'error' terms will always be available for this purpose. (In the SGD example the measured $J_z$ does not commute with the conserved $J_x$.)

The applicability of conservation laws to quantum measurements is a source of confusion and difficulty for those attempting to understand quantum mechanics. It would be worthwhile to present the results in this paper to students who are curious about the matter.